\documentclass[useAMS,usenatbib,singlespacing]{mn2e}
\usepackage{graphicx,amssymb}
\title[Cosmic ray acceleration: spectral steepening ]
{Cosmic ray acceleration by shocks: spectral steepening due to turbulent magnetic field amplification}
\author[AR Bell, JH Matthews, KM Blundell ]
{AR Bell$^{1}$\thanks{E-mail:Tony.Bell@physics.ox.ac.uk},
JH Matthews$^{2}$, and KM Blundell$^{2}$\\
$^{1}$University of Oxford, Clarendon Laboratory, Parks Road, 
Oxford OX1 3PU, UK\\
$^{2}$University of Oxford, Astrophysics, Keble Road, 
Oxford OX1 3RH, UK\\
}
\begin{document}
\date{}
\pagerange{\pageref{firstpage}--\pageref{lastpage}} \pubyear{2019}
\maketitle
\label{firstpage}
\begin{abstract}
We show that the energy required to turbulently amplify magnetic field during cosmic ray (CR) acceleration
by shocks
extracts energy from the CR and steepens the CR energy spectrum.
\end{abstract}
\section {Introduction}
In its simplest form, the theory of  
particle acceleration by  shocks  predicts a power-law number  spectrum 
$n(p) \propto p^{-s }$, where $p$ is the magnitude of the cosmic ray (CR) momentum.
This is equivalent to a momentum space
distribution function $  f(p) \propto p^{-\gamma }$ with $\gamma =s+2$. 
Theory predicts $s=2$, $\gamma =4$, for a strong shock.
The measured spectrum of Galactic CR arriving at the Earth 
approximates to a power-law but with a steeper spectrum, $s \approx 2.7$.
The steepness of the spectrum is partly explained by high energy CR escaping more rapidly from the Galaxy,
but the underlying CR spectrum at source is still steeper than $s=2$,
probably around $s=2.36$ (Hillas 2005, 2006).
Similarly, the number spectra of synchroton-emitting electrons in the more powerful young supernova remnants (SNR) approximate to power-laws, 
but typically with $s>2$.

In this paper we show that the steepened spectrum may be the result of energy loss to turbulence and  magnetic field during CR acceleration.
Related discussions on the same topic may be found in Zirakashvili \& Ptuskin (2014) and Osipov et al (2019).
Both theory and observation indicate that the magnetic field at the outer shocks of young SNR is much larger
than the ambient field expected in the interstellar medium (ISM).
The magnetic field is amplified by the non-linear development of 
plasma instabilities driven by the
current of CR streaming ahead of the shock (Bell 2004, Matthews et al 2017).
The  growth of magnetic field and the associated turbulence extracts energy from the CR while they are being accelerated and consequently 
steepens the CR spectrum.
We derive the  spectral steepening  first by solving the Vlasov equation (Sections 2 to 4), and second by a simpler more intuitive calculation (Section 5).
In Sections 6 to 8 we consider the implications of our analysis for measured CR spectra,
and compare theory with observation in Section 9.
In Section 9, we also consider the relationship of this process to other processes that may steepen the spectrum of accelerated particles.

Because protons dominate the Galactic CR population for energies up to 100s TeV
we make the simplifying assumption in Sections 2 to 7 that CR are comprised entirely of protons.
In Sections 8 and 9 we relax this assumption and discuss the effect on CR electrons and
their observed synchrotron spectra.


\section { Vlasov formulation of shock acceleration}
First-order Fermi acceleration by shocks can be understood in a variety of equivalent ways.
Bell (1978a,b) calculates the mean energy gain each time a cosmic ray (CR) crosses the shock and derives the energy spectrum by balancing the energy gain against the probability of a CR escaping downstream.
In a different but equivalent formalism,
Krymskii (1977), Axford, Leer \& Skadron (1977) and Blandford \& Ostriker (1978) solve the transport equation for the CR distribution function $f(z,p)$ which is defined in the local fluid rest frame moving at velocity $u(z)$ at position $z$.
In each formalism the role of the electric field is disguised by the frame transformation between the different fluid velocities upstream and downstream of the shock.
In the magnetohydrodynamic (MHD) approximation there is zero electric field, ${\bf E}= - {\bf u} \times {\bf B}$,  in the local rest frame defined by the frame in which ${\bf u }=0$.
Since magnetic field deflects CR trajectories without changing the magnitude of momentum, the energy of a CR
is unchanging in the local rest frame.  In the above formulations of shock acceleration the CR energy is constant except when it crosses the shock.

Here we use a formalism, which is different again, in which CR acceleration is seen to take place in the upstream CR precursor.
We analyse CR acceleration by solving the transport equation for a CR distribution function defined everywhere in the rest frame of the shock with no transformation applied at the shock.
The result is the same as in previous analyses but the role of the electric field is more clearly delineated.
Moreover, for the purposes of this paper, it facilitates a more straightforward
inclusion of energy transfer between CR and the turbulence driven by CR currents upstream of the shock.

CR proton acceleration can be described by the Vlasov equation in six-dimensional ${\bf p},{\bf r}$ space:
$$
\frac {\partial f}{\partial t}+{\bf v}. \frac {\partial f}{\partial {\bf r}}
+e\left ( {\bf E}+{\bf v}\times {\bf B} \right ) . \frac {\partial f}{\partial {\bf p}}=0
\eqno(1)$$
where ${\bf v}=c{\bf p}/p$, $p=|{\bf p}|$,
and the distribution function $f$ is defined in the rest frame of the shock.
The MHD fluid velocity ${\bf u}$ does not appear explicitly in equation (1) but is implicitly present 
through the electric field,
${\bf E}= - {\bf u} \times {\bf B}$.
$f({\bf p},{\bf r})$ can be expressed as a tensor expansion (Johnston 1960) in which the first three terms are
$$
f({\bf p},{\bf r})= f_0 (p)+{\bf f}_1(p). \frac {\bf p}{p}
+{\bf f}_2(p):\frac {\bf p}{p}\ \frac {\bf p}{p} \   .
\eqno(2)$$
The first term on the right-hand side of equation (2) represents the isotropic part of the distribution.
The second term represents transport by diffusion and advection of $f_0$.
The third term represents the pressure tensor and transport of ${\bf f}_1$.
${\bf f}_1$ is a vector, and ${\bf f}_2$ can be expressed mathematically as a $3\times 3$ matrix.
As shown by Johnston (1960), the first two equations in the tensor expansion of the Vlasov equation are:
$$
\frac {\partial f_0}{\partial t}
+ \frac {e{\bf E}}{3} . \frac{1}{p^2} \frac {\partial (p^2  {\bf f}_1)}{\partial p}
+\frac {c}{3} \frac {\partial }{\partial {\bf r}}. {\bf f}_1
=0
\hskip 2 cm 
\eqno(3)$$
$$
\frac {\partial {\bf f}_1}{\partial t}
+c \frac {\partial f_0}{\partial {\bf r}}
+     e{\bf E}  \frac {\partial f_0}{\partial p}
+\frac {ce{\bf B}\times {\bf f}_1}{p} 
\hskip 3 cm
$$
$$
\hskip 1 cm
+\frac {2c}{5} \frac {\partial }{\partial {\bf r}}. {\bf f}_2
+\frac {2}{5}     e{\bf E} . \frac{1}{p^3} \frac {\partial (p^3  {\bf f}_2)}{\partial p}
=0
\eqno(4)$$
where the first equation expresses number conservation, and the second is the momentum equation.
Using  ${\bf E}= - {\bf u} \times {\bf B}$, equation (3) becomes
$$
\frac {\partial f_0}{\partial t}
+\frac {c}{3} \frac {\partial }{\partial {\bf r}}. {\bf f}_1
-\frac {1}{3}\frac{1}{p^2} \frac {\partial }{\partial p}
\left [
p^2 {\bf u}.(e{\bf B} \times {\bf f}_1 )
\right ]
=0 \  .
\eqno(5)$$
Since ${\bf u}.{\bf E}=0$, the scalar product of equation (4) with ${\bf u}$ gives
$$
\frac {\bf u}{c} .\frac {\partial {\bf f}_1}{\partial t}
 + {\bf u} .\frac {\partial f_0}{\partial {\bf r}}
+\frac {{\bf u} .(e{\bf B}\times {\bf f}_1)}{p} 
+\frac {2}{5} {\bf u} . \left ( \frac {\partial }{\partial {\bf r}}. {\bf f}_2 \right )
=0 \  .
\eqno(6)$$
The standard theory of first-order Fermi acceleration by non-relativistic quasi-parallel shocks usually makes the assumption that CR
transport is diffusive with CR scattered isotropically  by
fluctuations in the magnetic field that are stationary in the local fluid frame.
This allows  $\partial {\bf f}_1/\partial t$
and terms including  ${\bf f}_2$ to be neglected as small quantities in an expansion in terms of small $u/c$ and small $\lambda/L$,
where $\lambda$ is the CR scattering mean free path and
$L$ is the characteristic hydrodynamic scalelength.

Here we side-step the  diffusion description by using equation (6) to model transport.
CR scattering occurs through a correlation between ${\bf B}$ and the transverse component of ${\bf f}_1$
that is produced by CR streaming through the fluctuating magnetic field.
The time or spatial average of
$e{\bf B} \times {\bf f}_1 $ is non-zero and represents a force on CR.
The scalar product of the average of
$e{\bf B} \times {\bf f}_1 $ with  ${\bf u}$ passes energy from the macroscopic background plasma flow
and results in first-order Fermi shock acceleration.

Bell et al (2013) went beyond diffusion theory by solving the equations for the CR distribution in a magnetic field
generated by CR currents and modelled by MHD.
${\bf f}_2$ had to be included in the kinetic equation for CR
because ${\bf f}_2$ plays the essential role of transporting ${\bf f}_1$ currents.
The inclusion of ${\bf f}_2$ allows for the excitation of turbulence 
through the non-resonant hybrid (NRH) instability (Bell 2004),
the consequent loss of CR energy to turbulence, and also the possibility of CR energy gain from turbulence
through second-order Fermi processes. 
However, the inclusion of ${\bf f}_2$  meant that a third equation in the expansion was needed
to represent the generation of ${\bf f}_2$  by gradients in ${\bf f}_1$.
Here we avoid the
addition of the  equation for $\partial {\bf f}_2/ \partial t$ by rewriting equation (4) to include a term
expressing energy exchange between CR and the turbulence. 

In this paper we  separate the motion of the background fluid into two parts 
$$
{\bf u}={\bf u}_h+{\bf u}_L
\eqno(7)
$$
defined as,
(i) the large scale hydrodynamic motion with velocity ${\bf u}_h$ of plasma flowing into the shock, 
changing discontinuously
at the shock, and flowing away downstream of the shock,
(ii) turbulent motion moving with velocity ${\bf u}_L$ on the scale of the Larmor radius $r_g=p/eB$ of the CR,
stretching and amplifying the magnetic field which in turn scatters the CR.

The ${\bf u}.(e{\bf B} \times {\bf f}_1 )$ term in equation (5) is split into two parts by
separating ${\bf u}$ into ${\bf  u}_h$ and ${\bf u}_L$.
CR energy gain from first-order diffusive shock acceleration is represented by ${\bf u}_h.(e{\bf B} \times {\bf f}_1 )$.
An  additional energy exchange between CR and turbulence is represented 
in equation (5) by ${\bf u}_L.(e{\bf B} \times {\bf f}_1 )$.

An expression for ${\bf u}_h.(e{\bf B} \times {\bf f}_1 )$ in equation (5) can be derived from 
equation (6)  by considering the components of equation (6) on the hydrodynamic scale
and averaging over the Larmor scale.
$\partial {\bf f}_1/\partial t$
and ${\bf f}_2$ in equation (6) can be neglected on the large scale since they average to zero
when integrated over hydrodynamical distance scales $\sim (c/u) r_g$ and over 
timescales for first-order acceleration. 
On the hydrodynamic scale, equation (6) averages to
$$
{\bf u}_h.(e{\bf B}\times {\bf f}_1)
=
-p{\bf u}_h. \frac {\partial f_0}{\partial {\bf r}} \  .
\eqno(8)
$$

The derivation of an expression for  ${\bf u}_L.(e{\bf B} \times {\bf f}_1 )$
is less straightforward.
On the smaller Larmor scale, ${\bf f}_2$ cannot be neglected.  
Closing the equations on the small scale would need the additional equation for 
$\partial {\bf f}_2/\partial t$.
Instead, we deal with the
${\bf u}_L.(e{\bf B} \times {\bf f}_1 )$ term in a different manner as described  in the next section. 

The overall equation for the CR distribution function is
$$
\frac {\partial f_0}{\partial t}
+
\frac {c}{3} \frac {\partial }{\partial {\bf r}}. {\bf f}_1
+
\frac{1}{3p^2} \frac {\partial }{\partial p}
\left (
p^3
 {\bf u}_h. \frac {\partial f_0}{\partial {\bf r}}
\right )
\hskip 2 cm 
$$
$$
\hskip 2 cm 
-  \frac{1}{3p^2} \frac {\partial }{\partial p}
\left [
p^2 {\bf u}_L.(e{\bf B} \times {\bf f}_1 )
\right ]
= 0 \  .
\eqno(9)$$
Solution of the equations without the fourth term gives first-order Fermi shock acceleration.
The fourth term represents the additional energy exchange between CR and turbulence.


\section { Energy loss to turbulence}
In this section we analyse the way in which ${\bf u}_L.(e{\bf B} \times {\bf f}_1 ) $ in equation (9) passes
energy from the CR to the turbulence.
$-e{\bf B} \times {\bf f}_1 $ represents a force pushing against
the turbulent plasma as it moves at velocity ${\bf u}_L$.
We may therefore approach this term from the perspective of its action on the  plasma turbulence 
as described by the equations
$$
\rho \frac {\partial {\bf u}}{\partial t}+ \rho ( {\bf u}.\nabla ) {\bf u}
= {\bf j} \times {\bf B}
\hskip 1. cm
\frac {\partial {\bf B}}{\partial t}=  \nabla \times ({\bf u} \times {\bf B})
$$
$$
\frac {\nabla \times {\bf B}}  {\mu _0} ={\bf j}  + {\bf j}_{CR}- n_{CR}e {\bf u}
\eqno(10)$$
where ${\bf u}$ is the plasma velocity defined in the shock rest frame as in Section 2,
${\bf j}_{CR}$ is the current density carried by the CR in the shock rest frame, and
${\bf j}$ is the current density carried by the background plasma in its local rest frame.
The final term $- n_{CR}e {\bf u}$ represents the current carried by the background plasma
with a charge density neutralising that of the CR which for simplicity we assume to consist purely of protons.
$- n_{CR}e {\bf u}$ is negligible when the growth of turbulence is analysed in the rest frame of the upstream plasma,
but is non-negligible when analysed in the shock rest frame.
Inclusion of $-\nabla P$ 
in the momentum equation, where $P$ is the thermal plasma pressure in the background plasma,
would add unnecessary complication without changing the essential point that energy is transferred to the turbulence through the $-{\bf j}_{CR} \times {\bf B}$ force operating on a small scale. 
The thermal pressure is initially small upstream of a strong shock, and only becomes significant upstream when the turbulence is well developed.

Energy transfer between CR and the background plasma can be found from equations (10)  by taking the scalar products of the momentum
equation with ${\bf u}$ and the induction equation with ${\bf B}/\mu _0$.
The rate at which the sum of the magnetic and kinetic energies changes is given by:
$$
\frac {\partial }{\partial t} \left (   \frac {\rho u^2}{2}   +\frac { B^2}{2 \mu _0}      \right )
+\nabla .  \left 
( \frac {{\bf E}\times {\bf B}}{\mu _0}
+  \frac { \rho u^2 {\bf u} }{2}     \right )
+{\bf j}.{\bf E}
= 
\hskip 1.5 cm
$$
$$
\hskip 3.5 cm
-{\bf u}. ({\bf j}_{CR} \times {\bf B})
\eqno(11)$$
where ${\bf E}\times {\bf B}/\mu _0$ is the Poynting flux,
and ${\bf j}.{\bf E}$ represents the rate of energy exchange with the thermal particles.
The right-hand side of the equation represents energy exchange with CR since  $-{\bf j}_{CR} \times {\bf B}$
is the force exerted by CR on the thermal plasma
as previously discussed by Zirakashvili, Ptuskin \& V\"{o}lk (2008).

By writing ${\bf u}={\bf u}_h+{\bf u}_L$   (equation (7))
we can separate the right hand side of equation (11) into two terms on the hydrodynamic and Larmor scales respectively.
The term $-{\bf u}_h. ({\bf j}_{CR} \times {\bf B})$ on the hydrodynamic scale describes the slowing of the plasma by the CR pressure as it flows into the shock; this term
extracts the energy needed to drive first-order CR acceleration.
The term $-{\bf u}_L. ({\bf j}_{CR} \times {\bf B})$  describes the rate at which energy 
is expended by CR on exciting the turbulence on the Larmor scale.

Since
$$
{\bf j}_{CR}=
\int _0^\infty
\frac {4\pi p^2ec}{3} {\bf f}_{1} dp
$$
and $ {\partial U _{turb}}/{\partial t}= -{\bf u}_L. ({\bf j}_{CR} \times {\bf B})$ is 
the local rate of energy transfer of energy to turbulence from CR,
$$
\frac {\partial U _{turb}}{\partial t}= \int _0 ^\infty \frac {\partial U_p}{\partial t} (p,{\bf r}) dp
$$
$${\rm where } \hskip .5 cm
 \frac {\partial U_p}{\partial t} (p,{\bf r})= \frac {4\pi p^2ec}{3} {\bf u}_L . ( {\bf B}\times{\bf f}_{1} )
\  ,
\eqno(12)$$
and
$({\partial U_p}/{\partial t} )dp$
is the rate at which energy is given to turbulence by CR with momenta
in the range $p$ to $p+dp$ at position ${\bf r}$.

From equations (9) and (12),
$$
 \frac {\partial f_0}{\partial t}
+
 \frac {c}{3}\frac {\partial }{\partial {\bf r}}. {\bf f}_1
+
\frac{1}{3p^2} \frac {\partial }{\partial p}
\left (
p^3
{\bf u}_h. \frac {\partial f_0}{\partial {\bf r}}
\right )
\hskip 1.5 cm
$$
$$
\hskip 1 cm 
- \frac {1}{4 \pi}\frac{1}{cp^2} \frac {\partial }{\partial p}
\left ( \frac {\partial\bar { U}_p}{\partial t}(p,{\bf r})
\right )
=0.
\eqno(13)$$
Equation (13) differs from equation (9) in that the fourth term is
now expressed in terms of energy gained by turbulence instead of energy lost by CR.


\section{ The CR spectrum}
We now derive the CR spectrum by spatially integrating equation (13) across the whole shock environment.
Far downstream of the shock, the CR distribution is isotropic in the fluid rest frame.
Correspondingly,
$f_1=-(u_\infty/c)p \partial f_0/\partial p$ in the rest frame of the shock
where $u_\infty$ is the fluid velocity far downstream.
Integration in $z$, parallel to the shock normal,
across the whole system and averaging in time such that $\partial f_0/\partial t$ can be omitted, gives
$$
u_\infty f_\infty
=
\frac {1}{3}
\int _ {-\infty}^{\infty}
\frac {\partial u}{\partial z}
\frac{1}{p^2} \frac {\partial }{\partial p}
\left (
p^3 f_0
\right ) dz
\hskip 2.5 cm 
$$
$$
\hskip 2.5 cm
+
\frac {1}{4 \pi c}
\frac {1}{p^2} \frac {\partial}{\partial p}
\left (
\int _{-\infty} ^\infty  \frac {\partial \bar {U}_p}{\partial t} (p,z) \  dz \right ).
\eqno(14)
$$
For simplicity, we neglect non-linear CR pressure feedback onto the hydrodynamics of the shock structure (Drury \& V\"{o}lk 1981).
With this simplification, the time average of 
$\partial u/\partial z$ is non-zero only at the shock and 
$f_0$ is uniform in the downstream plasma such that its value $f_s$ is the same as $f_\infty$.
This gives
$$
\left ( u_s-u_\infty\right ) p \frac {\partial  f_s}{\partial p} + {3u_ s}   f_s
=
\frac {3}{4 \pi p^ 2c}
\frac {\partial}{\partial p}
\left (
\int _{-\infty} ^\infty  \frac {\partial \bar {U}_p}{\partial t} (p,z) \  dz \right )
\eqno(15)$$
where $u_s$ is the shock velocity.
The integral on the right-hand side of equation (15) can be simplified using the 
continuity equation for the conservation of turbulent energy.
$ {\partial \bar {U}_p}/{\partial t} $ is small downstream of the shock so the integrand can be assumed to be non-zero
only upstream of the shock, giving
$$
\int _{-\infty} ^\infty  \frac {\partial \bar {U}_p}{\partial t} (p,z) \  dz 
= u_s \bar {U}_{ps} \  .
\eqno(16)$$
$u_s  \bar {U}_{ps}(p) dp$ is the rate at which energy is  given to turbulence by CR
where $ \bar {U}_{ps}(p) dp$ is the turbulent energy density immediately upstream of the shock
in the momentum range $p$ to $p+dp$.

We introduce $\Phi (p)$ to represent the 
ratio of the turbulence energy density immediately upstream of the shock to the CR
energy density at the shock.
$\Phi (p)$ is defined by
$ \bar {U}_{ps}(p) dp = 4 \pi p^3 c \Phi f_s(p) dp$.
If $\Phi$  is a slowly varying function of $p$, that is, $p|d\Phi/dp |\ll \Phi$, then 
$$
\left ( u_s-u_\infty -3 \Phi u_s \right ) p \frac {\partial  f_s}{\partial p} + \left ( 1 -3 \Phi \right ) {3u_ s}  f_s=0
\eqno(17)$$
giving
$$
f_s \propto p^{-\gamma} 
\hskip 0.3 cm {\rm where} \hskip 0.3 cm
\gamma = \frac {3u_s-9\Phi u_s}{u_s-u_\infty - 3 \Phi u_s}\  .
\eqno(18)
$$
In the limit of $\Phi \rightarrow 0$, the power-law index simplifies
to the well-established expression for diffusive shock acceleration in the absence of energy loss.
For strong shocks ($u_\infty = u_s/4$),
$$
\gamma = \frac {4(1-3 \Phi )}{1- 4\Phi}\ .
\eqno(19)
$$
In the limit of a strong shock and small $\Phi$, $\gamma = 4+4\Phi $.

The turbulence undergoes compression at the shock, but
the energy required to compress the turbulence 
is extracted from the large-scale hydrodynamic flow, not
from the CR.
If the turbulence is compressed adiabatically at the shock with an adiabatic index $\Gamma$,
where $\Gamma$ depends on the relative fractions of the magnetic, kinetic, and thermal energy density and
on whether the turbulence is isotropic, then the turbulence energy density far downstream  is
$$
\bar {U}_{p\infty} (p)= \left ( \frac {u_s}{u_\infty} \right )^{\Gamma} \bar {U}_{ps}(p)
= 4 \pi p^3 c \Phi _\infty f_s(p)
$$
$${\rm where} \hskip 0.5 cm
\Phi_\infty = \left ( \frac {u_s}{u_\infty} \right )^{\Gamma} \Phi \  ,
\eqno(20) $$
enabling equation (19) for the strong shock limit to be rewritten as
 $$
\gamma = 4 \left ( \frac { 4 ^\Gamma- 3\Phi _\infty   }
{ 4^{\Gamma }-  4 \Phi _\infty }  \right ) \ .
\eqno(21)
$$
The CR spectral  index approximates to $\gamma=4+4^{1- \Gamma} \Phi _\infty$ in the limit of small $\Phi _\infty $.
The difference between  $\bar {U}_{ps}$ defined immediately upstream of the shock 
and $\bar {U}_{p\infty} $ defined far downstream may be important when
comparing theoretical predictions with observations of synchrotron emission in the far downstream plasma.

\section { A heuristic derivation of the CR  spectrum}
Equation (18) for the CR spectrum  can be derived more simply, avoiding microphysical details, as follows.
Let $n(p)dp$ be the number density of CR at the shock with momenta in the range $p$ to $p+dp$.
The  rate at which CR cross the shock from upstream to downstream is $nc/4$.
The CR energy density at the shock in the  range $p$ to $p+dp$
is $ncp dp$.
By definition of $\Phi$, the turbulent energy density at the shock is  $\Phi n cp$,
and the rate at which turbulent energy is carried into unit area of the shock is $\Phi n cp u_s$.
Hence the energy $\Delta E_{\rm turb}$ lost to turbulence per shock crossing by a CR with energy $E=cp$ is $\Phi nu_s E$ divided by $nc/4$, giving 
$$
\Delta E_{\rm turb}= \frac {4 \Phi  u_s E}{c} .
\eqno(22)$$
According to Bell (1978a,b), the energy gained by CR at each crossing is
$$
\Delta E_{\rm accel}= \frac {4}{3} \frac {u_s- u_\infty}{c}E
\eqno(23)$$
The net energy gain per CR per crossing is
$$
\Delta E= \Delta E_{\rm accel}-\Delta E_{\rm turb}.
\eqno(24)$$
From Bell (1978a,b), the fraction by number of CR lost downstream between each shock crossing is
$$
\frac {\Delta N}{N}=-\frac {4u_\infty}{c}
\eqno(25)$$
where $N(p)=\int _p^\infty n(p)dp$.
Hence the integrated CR energy spectrum is
$$
\frac {d N}{d E} \approx 
\frac {\Delta N}{\Delta E}=- \frac {3 u_\infty}{u_s-u_\infty -3 \Phi u_s}
\ \frac {N}{E} \   .
\eqno(26)$$
The corresponding differential spectral index,
$\gamma = 3 - ( {E}/{N}) {dN}/{dE} $,
of the power-law for $f_s$ in momentum space is
$$
\gamma =
\frac {3u_s-9\Phi u_s }{u_s-u_\infty -3 \Phi u_s}
\eqno(27)$$
in agreement with equation (18).
\section {The shape of the CR spectrum}
The growth of turbulence and the amplification of magnetic field is driven by electric currents carried by CR
in the shock precursor.
CR are more numerous at low energies so they carry a larger current density.
If the turbulence and magnetic field amplification were driven only by low energy CR
the CR energy spectrum  produced by a shock would be steepened 
at low CR energy, but remain unaffected at high CR energy.
The result would be a concave CR energy spectrum resembling that predicted for non-linear feedback
(Drury 1983, Bell 1987, Blandford \& Eichler 1987,  Falle \& Giddings 1987, Jones \& Ellison 1991, Reynolds \& Ellison 1992).

However, the lower current density carried by high energy CR is compensated for by the larger distance their precursor extends upstream.
The following argument indicates that Galactic CR 
at high as well as low energy 
efficiently generate turbulence
and that the spectrum of CR accelerated by young SNR is thereby steepened over its full range.

As shown by Lagage \& Cesarsky (1983a,b), CR acceleration by SNR falls
far short of the knee at a few PeV in the Galactic CR spectrum unless
the magnetic field is amplified by a factor of $\sim 100$ beyond typical ambient interstellar values.
For CR at PeV energies to be scattered effectively, a component of the amplified magnetic field must be structured on the 
Larmor scale of PeV CR.
CR amplify magnetic field, probably by the non-linear NRH instability, on scales up to, but not greater than, their Larmor radius.
Hence the component of the magnetic field that scatters PeV CR must be generated by CR with a similar high energy,
and CR at high energies, as well as low energies, must play a role in amplifying the magnetic field.

Furthermore, the CR scattering mean free path, and the distance CR diffuse ahead of the shock,
is proportional to their Larmor radius.
Hence, low energy CR are unable 
to generate turbulence and magnetic field far enough ahead of the shock to scatter the highest energy CR.
This indicates that CR right up to the highest energies must drive turbulence,
and CR energy loss to turbulence is not limited to the more populous low momentum CR.
Hence $\Phi$ is large over the full energy range of CR acceleration,
and spectral steepening occurs over the whole momentum spectrum.

The argument derived from observation in this section is supported by more detailed theory in the next section.

\section {An estimate for $\Phi$}
The value of $\Phi$ can be estimated from the theory of the non-linear NRH instability
that is thought to be responsible for magnetic field amplification.
A key step in our derivation is the argument that
if turbulence and the amplified magnetic field are represented by a spectrum in wavenumber $k$,
then CR with Larmor radius $r_g$ excite and are
predominantly scattered
by magnetic field with a characteristic wavenumber $k\sim 1/r_g$.
The following derivation builds on a discussion to be found in Bell (2004).

The linear growth rate of the NRH instability is proportional to $\sqrt{k}$, so the small scale, large wavenumber, modes grow more quickly.
The initial largest wavenumber at which the instability grows is set by the requirement that
the ${\bf j}_{CR}\times {\bf B}$ force exerted by CR on the background plasma
must exceed the magnetic tension ${\bf B}\times (\nabla \times {\bf B})/\mu _0$ 
which opposes the stretching and amplification of magnetic field.
The wavenumber of the fastest growing mode is therefore $k_{\max}\approx \mu _0 j_{CR}/B$.

The instability grows non-linearly by the expansion and stretching of loops of magnetic field (Matthews et al 2017),
and the effective wavenumber decreases as the magnetic field is amplified.
Non-linear unstable growth continues, and the dominant wavenumber decreases, until a minimum wavenumber $k_{\min}$ is reached.
$k_{\min} $ is set by the requirement  that, for growth, the CR driving the instability must not be tied to magnetic field lines
and that the CR Larmor radius should exceed $1/k$.
Hence $k_{\min}\approx eB/p$.
Non-linear growth amplifies the magnetic field by large factors, typically up to 100,  in young SNR.
Since $k_{\min} \propto B$ and $k_{\max}  \propto B^{-1}$ the range of unstable wavenumbers
shrinks until $k_{\max} \approx k_{\min}$.
Consequently, the instability stops growing and saturates  when $k_{\min}\approx k_{\max}\approx 1/r_g$:
$$
\mu _0 j_{CR}/B \approx eB/p
\hskip 0.2 cm {\rm which \ gives} \hskip 0.2 cm
\frac{B ^2}{2 \mu _0} \approx \frac {p j_{CR}}{2e}.
\eqno {(28)}$$
By the nature of exponential growth, most of the energy transfer to turbulence
occurs during the final e-folding.
Hence, most of the energy input to turbulence occurs as it approaches saturation
and CR couple most strongly to 
turbulence structures on the scale of the CR Larmor radius.
Not only are CR most strongly scattered by turbulence on the CR Larmor scale,
but also CR energy loss to turbulence occurs most strongly on the Larmor scale.
On this basis, we assume that the CR momentum spectrum can be treated as being divided into momentum bands
with $\Delta p/p \sim 1$ interacting with wavenumber bands $\Delta k/k \sim 1$ in the turbulence
such that $k r_g \sim 1$.

The CR number density in the momentum  range  $[p,p+dp]$ is $(U_{CR}/pc)dp$ where $U_{CR}=4 \pi p^3 c f_s(p)$.
In the rest frame of the upstream plasma, CR drift at the shock velocity $u_s$,
giving an electric current $j_{CR}=eu_s U_{CR}/pc$,
where $j_{CR}dp$ is the electric current carried by CR in the  range $[p,p+dp]$.

In Section (4) we introduced $U_p dp$ as the energy density of turbulence excited by 
CR in the range $[p,p+dp]$.
$U_p$ includes thermal, magnetic and kinetic energy densities.
We define $U_m$ as the corresponding magnetic energy density,
and $\epsilon=U_m/U_p$ as their ratio.
From equation (28) $U_m  dp\approx (p/2e) j_{CR}dp$, giving
$$
U_m\approx\frac {u_s}{2c} U_{CR}
\eqno {(29)}
$$
as supported by observation (Vink 2008).

Since $\Phi=U_p/U_{CR}$ and $U_m=\epsilon U_p$, we now have an estimate for $\Phi $:
$$
\Phi \approx \frac{1}{2 \epsilon}\ \frac {u_s}{c} .
\eqno{(30)}
$$
Note that $\Phi$, and therefore the spectral steepening, depends only on the ratio ($U_p/U_{CR}$)
of the energy density of the turbulence to the energy density
of the CR.
Spectral steepening occurs even if both of these are much smaller than the hydrodynamic energy density $\rho u_s^2$.
Although the spectral steepening discussed here is a non-linear effect in the sense that it arises from the growth of turbulence,
it is different from spectral steepening due to shock profile modification which arises from a large ratio of $U_{CR}$ to $\rho u_s^2$.

\section{The electron spectrum}
Equation (13) may be applied to electrons or any nuclei with $Z>1$,  as well as to protons, 
with $\partial U_p /\partial t$ representing the rate at which energy is given to the turbulence by the relevant species.
However, there is no guarantee that other CR species transfer energy to turbulence at the same rate
as protons.
Indeed, there is no guarantee that they lose energy to the turbulence at all.
A scenario is conceivable in which a majority species, most likely to be protons,
generates turbulence, and then minority species gain energy by second-order Fermi acceleration through $\partial U_p /\partial t$ 
 in their version of  equation (13).
By this process, the electron spectrum could be flattened, instead of steepened,
by interaction with the turbulence.

The direction of energy transfer between CR electrons and turbulence
depends on the detailed microphysics of the process.
One might argue that energy is transferred through the electric field, and that
an electron following the same trajectory through an electric field would gain energy
where a proton would lose energy due to its opposite charge.
In their equation (8), Zirakashvili, Ptuskin \& V\"{o}lk (2008) include the second-order electric field 
${\bf E}_0=- \delta {\bf u} \times \delta {\bf B}$
where $ \delta {\bf u}$  and $\delta {\bf B}$ are respectively 
the perturbed fluid velocity and perturbed magnetic field
in the linear analysis of the plasma instability.
As they show, ${\bf E}_0$ is a large-scale electric field aligned with the zeroth order CR current
such that 
${\bf E}_0$ extracts energy from the CR particles driving the instability
(Zirakashvili \& Ptuskin 2008, 
 Zirakashvili, Ptuskin \& V\"{o}lk 2008, Osipov et al 2019).
It follows that if CR protons lose energy to drive the instability, then CR electrons should gain energy from 
${\bf E}_0$.
However, when viewed from the rest frame of the shock,
${\bf E}_0$ generates an electric potential upstream of the shock with the consequence that
protons or electrons crossing the shock into the upstream region and then returning to the shock 
neither gain nor lose energy through ${\bf E}_0$ in the frame in which equation (13) is formulated.
Moreover, in the non-linear phase of fully-developed turbulence on the Larmor scale that dominates both magnetic field amplification and CR scattering, 
the structure of the  electric field is more complicated,
and CR with opposite charges are deflected in opposite directions with the resulting possibility
of all species gaining energy from a disordered electric field.

A more appropriate way of understanding fully non-linear energy exchange between CR and turbulence may follow from 
the interaction between the CR pressure gradient and Alfv$\acute{{\rm e}}$n turbulence as considered by Wentzel (1974), Skilling (1975a,b,c)
and subsequent authors.
The process is similar to that described by
the third term of equation (13) which represents CR energy gain due to ${\bf u}_h.\nabla P_{cr}$
and arises from work done by hydrodynamic  flow against the CR pressure $P_{cr}$.
Both protons and electrons gain energy through  ${\bf u}_h.\nabla P_{cr}$,
even though they have different charges, since the protons and electrons follow different trajectories
through the turbulence.
A similar term ${\bf u}_t .\nabla P_{cr}$ can describe energy transfer between CR and turbulence,
where ${\bf u}_t $ is the velocity of the turbulent magnetic field relative to the background fluid
(eg Malkov \& Drury 2001, Bell \& Lucek 2001).
If the turbulence were to  consist of linear Alfv$\acute{{\rm e}}$n waves, then the magnitude of ${\bf u}_t$ is characteristically 
the Alfv$\acute{{\rm e}}$n speed at which the waves move.
In our case the turbulence is non-linear and strongly driven, so
 $|{\bf u}_t|$ might depart considerably from the Alfv$\acute{{\rm e}}$n speed.
Nevertheless, the same principle may hold.
Since  ${\bf u}_t$ is on average the same for both electrons and protons,
a case can be made that both electrons and protons lose energy to turbulence
and at a similar rate.
If so, the electron and proton spectra can be expected to be steepened to the same degree.
A scenario in which both electrons and protons drive, and lose energy to, the turbulence
is consistent with observations as discussed in the next section.

This matter requires further consideration, especially regarding the role of the second-order electric field 
(Zirakashvili, Ptuskin \& V\"{o}lk 2008,
 Zirakashvili \& Ptuskin 2014,
Osipov et al 2019)
relative to that of the disordered electric field in fully developed turbulence.
Such a discussion is beyond the scope of this paper.


\section{Application to observations}
CR arriving at the Earth have a spectral index $s \approx 2.7$ at CR energies 
up to the observed knee in the spectrum at a few PeV.
Preferential escape from the Galaxy during propagation at high CR energies
implies a  spectral index around 2.36 at source, corresponding to $\gamma \approx 4.36$
(Hillas 2005, 2006).
From equation (19),
a spectrum at source with $\gamma =4.36$ implies $\Phi \approx 0.066$
suggesting that of the order of $ 5-10\%$ of the CR energy is given to turbulence in the upstream plasma if the
CR are accelerated by a shock.

Synchrotron spectra provide further information on the spectra of CR electrons accelerated by shocks.
The spectral index $\alpha$ of synchroton emission is related to the CR electron spectral index by
$\alpha=(\gamma -3)/2$.
Generally, the radio synchrotron spectra of young SNR are
steeper than the spectral index $\alpha =0.5$  expected for electrons
accelerated at a strong shock.
The magnetic field is known to be strongly amplified at the outer shocks of
young SNR (Vink \& Laming 2003, V\"{o}lk et al 2005).
If the  argument in Section 8 is correct that both the electron and proton spectra are steepened to the same degree,
then, from equation (18) for a strong shock with $u_\infty= u_s/4$,
$$
\alpha = \frac {1}{2}\ \frac {1}{1 -4\Phi}.
\eqno {(31)}$$
From equation (30), the relation between $\Phi $ and the shock velocity depends on
the fraction $\epsilon$ of the turbulent energy density residing in magnetic field.
An obvious guess is that energy in the turbulence is shared equally between magnetic energy,
kinetic energy and thermal energy, giving $\epsilon \approx 1/3$,
in which case, equations (30) and (31) combine to give the dependence of the radio spectral index on the shock velocity:
$$
\alpha = \frac {1}{ 2 - 12 u_s/c}
\eqno {(32)}$$
Fig. 1 is a plot of the synchrotron spectral index $\alpha$ of several supernova remnants (SNR) 
against their expansion velocity
defined as the radius $R$ divided by the age $t$.
The data is taken from Table 1 and Fig. 4 of Bell et al (2011),
where information on  the identities of the SNR and the uncertainties in the data can be found.
The relationship in equation (32) between $\alpha$ and $u_s$ is plotted as the curve in Fig. 1.
The data show the expected increase in $\alpha$ at high shock velocities.

The CR spectrum can  be steepened by other factors, including
(a) if the CR pressure is large,
non-linear feedback can steepen the spectrum at low CR energy (see references in Section 6);
(b) at shock velocities approaching $c$, the shock compression deviates from the
non-relativistic compression and the CR distribution at the shock becomes anisotropic to high order 
(eg Heavens \& Drury 1988, Kirk et al 2000,  Achterberg et al 2001);
(c) the presence of high-order CR anisotropies at oblique and quasi-perpendicular shocks can steepen CR spectra
at shock velocities exceeding $\sim c/10$ (Bell et al 2011).
The process leading to spectral steepening discussed in this paper is distinct from those due
to non-linear feedback, relativistic effects, or shocks being quasi-perpendicular.

More than one of these factors may be important at the same time.
In particular, given the efficiency with which CR are produced in the Galaxy,
it would be surprising if non-linear feedback were negligible.
The expected signature of non-linear feedback is spectral steepening at low CR energy due to smoothing of the shock structure,
and spectral flattening at high energy caused by increased compression at the shock (see references listed in Section 6).
Reynolds \& Ellison (1992) find a hint of concave curvature in a study of the Tycho and Kepler SNR, but the evidence is not compelling.
In contrast, the spectrum of Cassiopeia A, which is both steep ($\alpha=0.75$, $s=2.5$) and straight from 25MHz to 250GHz (Vinyaikin 2014), points to a strong
steepening process that is equally present across two orders of magnitude in relativistic electron energy.

In some circumstances, non-linear shock modification and energy loss to turbulence might  play a mutually interacting role in shaping the CR energy  spectrum.
For example, if the spectrum is significantly steepened by energy loss to turbulence,
mildly relativistic protons dominate the CR pressure.
Spectral concavity due to non-linear shock modification would then  be confined to the mildly relativistic part of the CR spectrum,
and the rest of the spectrum would be straight.

\begin{figure}
\includegraphics[angle=0,width=8cm]{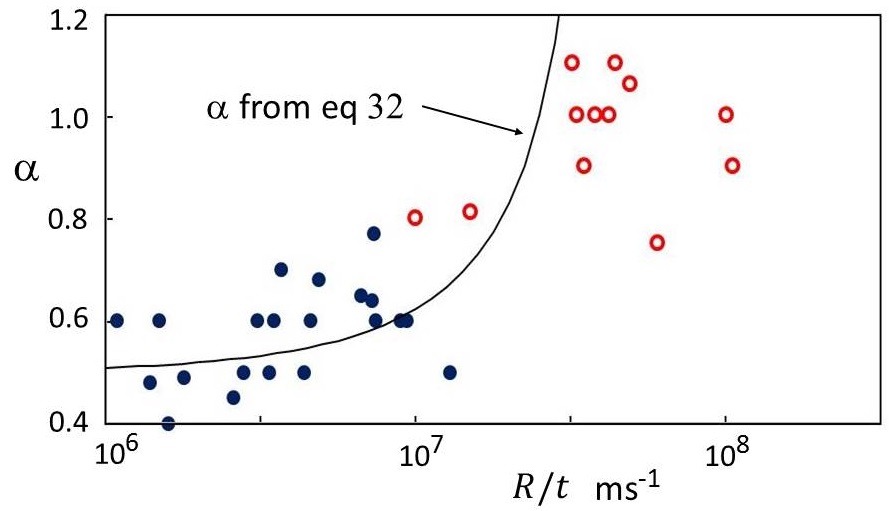}
\centering
\caption{
Plot of the radio spectral index $\alpha$ against the mean expansion velocity $R/t$
of Galactic (blue dots) and extragalactic (red open circles) SNR.
The data is taken from Bell et al (2011).
The curve is a plot of $\alpha$ as predicted by equation (32).
}
\label{fig:figure1}
\end{figure}


\section{Conclusions}
An unresolved theoretical question regarding particle acceleration by shocks is the contrast between the spectrum predicted by theory,
$n(p) \propto p^{-2}$, and
the steeper spectra of Galactic CR and synchrotron-emitting electrons in many radio sources.
The steepening of the Galactic CR extends from GeV to PeV energies.
Radio synchrotron spectra can be curved, but spectral steepening is often present across the whole radio spectrum.
We show that this  steepening  may be caused by the loss of CR energy to 
turbulence  and magnetic field.
Our analysis predicts, consistent with observation, that a greater fraction of the CR energy is lost to turbulence when the shock velocity is high, and consequently that the CR spectrum is generally steeper when CR are accelerated by shocks
with a high velocity.
The spectral index of Galactic CR is consistent with CR acceleration by young SNR.

Spectral steepening as considered here is non-linear in the sense  that it depends on the shock velocity
and on non-linear turbulent amplification of  magnetic field.
However, the spectral steepening does not depend on the ratio of the CR pressure to the kinetic pressure $\rho u_s^2$
at the shock,
and in that sense it is linear.

\section {ACKNOWLEDGEMENTS}
We thank Luke Drury, Anabella Araudo, Roger Blandford and Donald Ellison for interesting and helpful discussions and comments.
This research was supported by the UK
Science and Technology Facilities Council under grant No. ST/N000919/1.

\section{REFERENCES}
Achterberg A., Gallant Y.A., Kirk J.G., Guthmann A.W., 2001, MNRAS 328, 393
\newline
Axford W.I., Leer E., Skadron G., 1977,  Proc 15th Int. Cosmic Ray Conf., 11, 132
\newline
Bell A.R., 1978a, MNRAS, 182, 147 
\newline
Bell A.R., 1978b, MNRAS, 182, 443
\newline
Bell A.R., 1987, MNRAS, 225, 615
\newline
Bell A.R., 2004, MNRAS 353, 550
\newline
Bell A.R., Lucek S.G., 2001, MNRAS 321, 433
\newline
Bell A.R.,  Schure K.M., Reville B., 2011, MNRAS 418, 1208
\newline
Bell A.R., Schure K.M., Reville B., Giacinti G., 2013,  MNRAS 431, 415
\newline
Blandford R.D.,  Ostriker J.P.,  1978, ApJ 221, L29
\newline 
Blandford R.D., Eichler D., 1987, Phy Rep 154, 1
\newline
Drury L.O'C.,  V\"{o}lk H.J., 1981, ApJ 248, 344
\newline
Drury L.O'C., 1983, Rep Prog Phys 46, 973
\newline
Falle S.A.E.G, Giddings J.R., 1987, MNRAS 225, 399
\newline
Heavens A.F., Drury L.O'C., 1988, MNRAS 235, 997
\newline
Hillas A.M., 2005, J Phys G 31, R95
\newline
Hillas A.M., 2006, arXiv:astro-ph/0607109
\newline
Johnston T.W., 1960, Phys Rev 120, 1103
\newline
Jones F.C., Ellison D.C., 1991, Space Science Reviews 58, 259
\newline
Kirk J.G., Guthmann A.W., Gallant Y.A., Achterberg A., 2000, ApJ 543, 235
\newline
Krymskii G.F.,  1977, Sov Phys Dokl, 23, 327
\newline
Lagage P.O., Cesarsky C.J., 1983a, A\&A 118, 223
\newline
Lagage P.O., Cesarsky C.J., 1983b,  ApJ 125, 249
\newline
Malkov M.A., Drury L.O'C, 2001, Rep Prog Phys 64, 429
\newline
Matthews J.H., Bell A.R., Blundell K.M., Araudo A.T., 2017, MNRAS 469, 1849
\newline 
Osipov S.M.,  Bykov A.M., Ellison D.C., 2019, J Phys: Conf Series, in press, arXiv:1905.10266
\newline
Reynolds S.P., Ellison D.C., 1992, ApJ 399, L75
\newline
Skilling J., 1975a, MNRAS 172,557
\newline
Skilling J., 1975b, MNRAS 172, 245
\newline
Skilling J., 1975c, MNRAS 173, 255
\newline
Vink J., 2008, AIP proc 1085, 169
\newline
Vink J., Laming J.M., 2003, ApJ, 584, 758
\newline
Vinyaikin E.N., 2014, Astr Rep 58, 626
\newline 
V\"{o}lk H.J., Berezhko E.G., Ksenofontov L.T., 2005, A\&A 433, 229
\newline
Wentzel D.G., 1974, ARA\&A 12, 71
\newline
Zirakashvili V.N., Ptuskin V.S., 2008,  ApJ 678, 939
\newline
Zirakashvili V.N., Ptuskin V.S., 2014,  Bulletin of the Russian Academy of Science 79, 316
\newline
Zirakashvili V.N., Ptuskin V.S.,  V\"{o}lk H.J., 2008, ApJ 678, 255
\newline
\end{document}